\documentclass[prd,twocolumn,floatfix,amsmath,nofootinbib,amssymb,floatfix]{revtex4}
\usepackage{graphicx,color,dcolumn,booktabs,bm,multirow}
\usepackage{longtable,lscape}
\usepackage{txfonts}
\usepackage{overpic}
\usepackage{amssymb}
\usepackage{array}
\usepackage{indentfirst}
\usepackage{feynmf}   
\usepackage{bbding}
\usepackage{slashed}  
\usepackage{cases}
\usepackage{color}
\usepackage{multirow}
\usepackage{epstopdf}
\usepackage{tabularx}
\usepackage{graphicx,color,dcolumn,booktabs,bm}
\usepackage[compat=1.1.0]{tikz-feynman}
\usepackage[colorlinks,
citecolor=blue,
anchorcolor=red,
menucolor=red,
linkcolor=red,
filecolor=red,
runcolor=red,
urlcolor=blue,
frenchlinks=red]{hyperref}
\usepackage{float}

\begin{document}
\title{$\Lambda_{c}(2910)$ and $\Lambda_{c}(2940)$ productions in $\pi^{-} p$ scattering process}
\author{Quan-Yun Guo$^{1}$}
\author{Dian-Yong Chen$^{1,2}$\footnote{Corresponding author}}\email{chendy@seu.edu.cn}
\affiliation{$^1$ School of Physics, Southeast University, Nanjing 210094, People's Republic of China}
\affiliation{$^2$ Lanzhou Center for Theoretical Physics, Lanzhou University, Lanzhou 730000, China}
\date{\today}

\begin{abstract}
In the present work, we propose to investigate the productions of $\Lambda_{c}(2910)$ and $\Lambda_{c}(2940)$ in the $\pi^{-} p \rightarrow D^{-} D^{0} p$ processes by utilizing an effective Lagrangian approach, where $\Lambda_c(2910)$ and $\Lambda_c(2940)$ are considered as $D^\ast N$ molecular states with $J^P$ quantum numbers to be $1/2^-$ and $3/2^-$, respectively. With the cutoff parameter determined by the upper limit of the cross sections for $\pi^- p \to D^{\ast-} \Lambda_c(2286)$, the ratios of the cross sections for $\pi^- p \to D^{\ast-} \Lambda_c(2286)$, $\pi^- p \to D^{-} \Lambda_c(2286)$, $\pi^- p \to D^{-} \Lambda_c(2910)$, and $\pi^- p \to D^{-} \Lambda_c(2940)$ are estimated to be $1:4.8:1.42:0.26$ at $p_\pi=30$ GeV. Considering that the $\Lambda_{c}(2910)$ and $\Lambda_{c}(2940)$ state can further decay into $D^{0}p$, we estimate the cross sections for $\pi^{-} p \rightarrow D^{-} D^{0} p$ process and the differential cross sections depending on the $D^0 p$ invariant mass spectrum. Our estimations indicate that the total cross sections are $(0.49^{+1.56}_{-0.38})$ nb when $p_{\pi}=15~\mathrm{GeV}$, where the uncertainties result from the variation of the $\Lambda_{r}$. By comparing the contributions of the $s$, $u$, and $t$-channels, we conclude  that the $t$-channel plays the predominant role. Moreover, the present estimations suggest that the structure around 2.9 GeV in the $D^0 p$ invariant mass spectrum of the $\pi^{-} p \rightarrow D^{-} D^{0} p$ process should correspond to $\Lambda_c(2910)$ rather than $\Lambda_c(2940)$, which can be tested by further experimental measurements at J-PARC in the future.

\end{abstract}
	
\maketitle

\section{Introduction}
\label{sec:Introduction}
Over the past two decades, the exploration of multiquark candidates has emerged as a significant area of interest. Numerous new hadron states, potential exotic state candidates, have been identified by experimental Collaborations including Belle, BelleII, BESIII, and LHCb (See Refs.~\cite{Belle:2003nnu, Belle:2007hrb, Belle:2009and, BESIII:2013ris, Belle:2013yex, LHCb:2015yax, LHCb:2016axx, BESIII:2016bnd, LHCb:2020tqd, LHCb:2021vvq} for representative examples). Of these exotic candidates, the possible pentaquark states, $P_{c}(4380)$ and $P_c(4450)$, were initially observed in the $J/\psi p$ invariant mass distributions of the $\Lambda^{0}_{b} \rightarrow J/\psi K^{-}p$ process by the LHCb Collaboration in 2015~\cite{LHCb:2015yax,LHCb:2016ztz,LHCb:2016lve}. Subsequent analysis of the same process in 2019 suggested that the structure of $P_{c}(4380)$ could be described by a background effect, and a new narrow state, $P_{c}(4312)$, was observed for the first time, while $P_c(4450)$ appeared to split into two distinct states, $P_c(4440)$ and $P_c(4457)$~\cite{LHCb:2019kea}. Following these observations, the LHCb Collaboration further observed strange counterparts of the $P_c$ states, which are named the $P_{cs}$ states. The first $P_{cs}$ state, $P_{cs}(4459)$, was observed in the $J/\psi \Lambda$ invariant mass distributions of the $\Xi_b^- \to K^- J/\psi \Lambda $ process in 2020~\cite{LHCb:2020jpq}. Two years later, another state, $P_{cs}(4338)$, was reported in the $J/\psi \Lambda$ invariant mass spectrum of the decay $B^- \to J/\psi \Lambda \bar{p}$~\cite{LHCb:2022ogu}. The masses of the observed $P_{c}$ states are close to the thresholds of $D^{(\ast)} \Sigma_c$, while those of the $P_{cs}$ states are close to the thresholds of $D^{(\ast)}\Xi_c$, which leads to the prosperity of the molecular interpretations of $P_c$ and $P_{cs}$ states~\cite{Chen:2019bip, Chen:2019asm, Guo:2019fdo, Liu:2019tjn, Xiao:2019mvs, Xiao:2019aya, Zhang:2019xtu, Wang:2019hyc, Xu:2019zme, Burns:2019iih, Lin:2019qiv, He:2019rva, Du:2019pij, Wang:2019spc, Xu:2020gjl, Xu:2020flp, Ling:2021lmq, Chen:2021cfl, Du:2021bgb, Chen:2022onm, Ke:2023nra, Xu:2023ipd, Chen:2020kco, Chen:2021tip, Zhu:2021lhd, Xiao:2021rgp, Lu:2021irg, Wang:2022neq, Feijoo:2022rxf, Wu:2024lud,Wu:2019rog,Wu:2021caw}.

Beyond the extensively studied $P_c$ and $P_{cs}$ states, the charmed baryons $\Lambda_{c}(2910)$ and $\Lambda_{c}(2940)$ have also been suggested as potential pentaquark state candidates. Experimentally, $\Lambda_{c}(2940)$ was first observed in the $D^{0}p$ invariant mass distributions by the BABAR Collaboration using $287~\mathrm{fb}^{-1}$ of the annihilation data collected by the BABAR detector at a center-of-mass energy of 10.58 GeV in 2006~\cite{BaBar:2006itc}. Subsequently, the Belle Collaboration confirmed the observation of $\Lambda_{c}(2940)$ in the $\Sigma_{c}(2455)^{0,++} \pi^{+,-}$ invariant mass distributions, using a $553~\mathrm{fb}^{-1}$ data sample recorded by the Belle detector at or 60 MeV below the $\Upsilon(4S)$ resonance in 2007~\cite{Belle:2006xni}. In 2022, the Belle Collaboration discovered a new structure, $\Lambda_{c}(2910)$, in the $\Sigma_{c}(2455)^{0,++} \pi^{+,-}$ invariant mass spectrum of $\bar{B}^{0} \rightarrow \Sigma_{c}(2455)^{0,++} \pi^{+,-} \bar{p} $ processes with a significance of $4.2 \sigma$~\cite{Belle:2022hnm}. The measured resonance parameters of $\Lambda_{c}(2910)$ and $\Lambda_{c}(2940)$ are~\cite{ParticleDataGroup:2024cfk},
\begin{eqnarray}
\Lambda^{+}_{c}(2910) : &\mathrm{M}&=\left(2914 \pm 7 \right)\; \mathrm{MeV}, \nonumber\\ &\Gamma&=\left(52 \pm 27\right) \; \mathrm{MeV}, \nonumber\\ \Lambda^{+}_{c}(2940) : &\mathrm{M}&=\left(2939.6^{+1.3}_{-1.5}\right) \; \mathrm{MeV}, \nonumber\\ &\Gamma&=\left(20^{+6}_{-5}\right) \; \mathrm{MeV},
\end{eqnarray}
respectively.

Similarly to the case of $P_c$ and $P_{cs}$ states, the observed masses of $\Lambda_c(2910)$ and $\Lambda_c(2940)$ are close to the threshold of $D^\ast N$, which indicates that both $\Lambda_c(2910)$ and $\Lambda_c(2940)$ potentially are the $D^\ast N$ pentaquark molecular states. In Ref.~\cite{He:2006is}, the authors proposed that $\Lambda_{c}(2940)$ could be regarded as $D^{\ast 0} p$ molecular state with $J^{P}=1/2^{-}$. The estimations of the strong two-body decay channels $D^{0} p$, $\Sigma^{++}_{c} \pi^{-}$, $\Sigma^{0}_{c} \pi^{+}$ and three-body decay channels in Ref.~\cite{Dong:2009tg,Dong:2011ys} suggested that $\Lambda_c(2940)$ was consistent with the experimental data when $J^{P}=1/2^{+}$, which indicated that $\Lambda_c(2940)$ should be a $P$-wave $D^\ast N$ molecular state. Using the one-boson exchange model, the calculations of $D^{\ast} N$ bound state in Ref.~\cite{He:2010zq} indicated that $\Lambda_{c}(2940)$ could be explained as an isoscalar $S$-wave or an isoscalar $P$-wave state with $I(J^{P})=0(1/2^{+} $ or $ 0(3/2^{-})$. QCD sum rules were also applied to investigate the properties of $\Lambda_{c}(2940)$~\cite{Zhang:2012jk}, and the results showed that $\Lambda_{c}(2940)$ could be regarded as the $S$-wave $D^{\ast} N$ state with $J^{P}=3/2^{-}$ although there may be some computational limitations. Considering the mass similarity of the mass splittings of $P_c(4440)/P_c(4457)$ and $\Lambda_c(2910)/\Lambda_c(2940)$, the authors in Ref.~\cite{Yue:2024paz} investigated the decay properties of $\Lambda_c(2910)$ and $\Lambda_c(2940)$ in the $D^\ast N$ molecular frame by an effective Lagrangian approach, and the estimations suggested $\Lambda_c(2910)$ and $\Lambda_c(2940)$ could be assigned as $D^\ast N$ molecular states with $J^P=1/2^-$ and $3/2^-$, respectively. With the chiral effective field theory to the next-to-leading order, the estimations in Ref.~\cite{Wang:2020dhf} suggested that $\Lambda_c(2940)$ could be $D^\ast N$ molecular state, and the observed signal of $\Lambda_c(2940)$ might contain the spin$-1/2$ and spin$-3/2$ structures. The QCD sum rule calculations by carrying out the operator product expansion up to the vacuum condensates of dimension 13 and taking full account of the light-flavor SU(3) breaking effects indicated that $\Lambda_c(2940)/\Lambda_c(2910)$ could be $D^\ast N$ molecular state with the $3/2^-$~\cite{Xin:2023gkf}. In the constituent quark model, the decay properties of $\Lambda_c(2940)$ were investigated in the $D^\ast N$ molecular frame. By using the quark delocalization color screening model, the authors in Ref.~\cite{Yan:2022nxp} found that $\Lambda_c(2940)$ more likely to be a $D^\ast N$ molecular state with $J^P=3/2^-$, while $\Lambda_c(2910)$ could not be interpreted as a $D^\ast N$ molecular state.

In addition to the mass spectrum and decay properties, researches on the productions of $\Lambda_{c}(2910)$ and $\Lambda_c(2940)$ has also been performed. In Ref.~\cite{He:2011jp}, the cross sections of the $p \bar{p} \rightarrow \bar{\Lambda}_{c} \Lambda_{c}(2940)$ process were estimated, where the $J^{P}$ quantum numbers of $\Lambda_c(2940)$ were considered to be $1/2^{\pm}$, $3/2^{\pm}$ or $5/2^{\pm}$, and the magnitudes of the estimated cross sections were between $10^{0}$ and $10^{5}$ $\mathrm{nb}$. Similar processes were investigated in Ref.~\cite{Dong:2014ksa}, where $\Lambda_{c}(2940)$ was regarded as a $D^{\ast 0} p$ molecular state with $J^{P}=1/2^{\pm}$. In addition, Ref.~\cite{Wang:2015rda} proposed that the production of $\Lambda_{c}(2940)$ could be studied in the $\gamma n \rightarrow D^{-} \Lambda_{c}(2940)$ process.

Recently, the J-PARC hadron facility~\cite{Aoki:2021cqa} proposes that the expected pion energy will reach over 20 $\mathrm{GeV}$ in the laboratory frame, providing some new platform to investigate the productions of $\Lambda_{c}$ states. In Ref.~\cite{Xie:2015zga}, the authors proposed that the $\pi^{-} p \rightarrow D^{-} D^{0} p$ reaction could be investigated by the $t$-channel $D^{\ast 0}$ meson exchange, $s$-channel nucleon pole, and $u$-channel $\Sigma^{++}_{c}$ exchange, where $\Lambda_{c}(2940)$ was considered as a $D^{\ast 0} p$ molecular state with $J^{P}=1/2^{+}, 1/2^{-}$, respectively. The results indicated that in the above two cases, the $t$-channel provides the main contribution to the total cross section. After the observation of $\Lambda_{c}(2910)$ by Belle Collaboration in 2022, both $\Lambda_{c}(2910)$ and $\Lambda_{c}(2940)$ could be explained as $S$-wave $N D^{\ast}$ molecular states. In addition, the estimations in Ref.~\cite{Yue:2024paz} suggested that the $J^{P}$ quantum numbers of $\Lambda_{c}(2910)$ and $\Lambda_{c}(2940)$ are preferred to be $1/2^{-}$ and $3/2^{-}$, respectively. Therefore, both $\Lambda_{c}(2910)$ and $\Lambda_{c}(2940)$ are expected to contribute to the process $\pi^{-} p \rightarrow D^{-} D^{0} p$. It should be noted that the authors in Ref.~\cite{Christenson:1985ms} measured the cross sections for $\pi^{-} p \rightarrow D^{\ast-} \Lambda_{c}$ process at the AGS at Brookhaven National Laboratory, and the results showed that the upper limit of cross section is 7 nb when $P_{\pi}=13$ GeV, which can be employed to determine the model parameters involved in the present estimations. Thus, In the present work, we first estimate the cross sections for $\pi^{-} p \rightarrow D^{\ast-} \Lambda_{c}$,  with the experimental data in Ref.~\cite{Christenson:1985ms}, the model parameter $\Lambda_{r}$ can be determined. Then, with the same model parameter, the two body process $\pi^{-} p \rightarrow D^{-} \Lambda_{c} / \Lambda_{c}(2910) / \Lambda_{c}(2940)$ and the three body process $\pi^{-} p \rightarrow D^{-} D^{0} p$ are investigated by considering $\Lambda_{c}(2910)$ and $\Lambda_{c}(2940)$ to be $D^\ast N$ molecular states.  In addition, we also estimate the $D^0 p$ invariant mass distributions of the process $\pi^{-} p \rightarrow D^{-} D^{0} p$ to further check the contributions of $\Lambda_c(2910)$ and $\Lambda_c(2940)$.

This work is organized as follows. After introduction, we present our estimation of the cross sections for $\pi^{-} p \rightarrow D^{\ast-} \Lambda_{c}$, $\pi^{-} p \rightarrow D^{-} \Lambda_{c} / \Lambda_{c}(2910) / \Lambda_{c}(2940)$ and $\pi^{-} p \rightarrow D^{-} D^{0} p$ processes. In Section \ref{sec:MA}, the numerical results and related discussions of the cross sections are presented, and the last section is devoted to a short summary.

\section{$\Lambda_{c}(2286)/\Lambda_{c}(2910)/\Lambda_{c}(2940)$ productions in the $\pi^{-} p$ scattering process}
\label{sec:MS}

\subsection{Cross Sections for $\pi^{-} p \rightarrow D^{\ast -} \Lambda_{c}$ and $\pi^{-} p \rightarrow D^- \Lambda_{c} / \Lambda_{c}(2910) / \Lambda_{c}(2940)$}

\begin{figure*}[htb]
     \includegraphics[width=165mm]{TwobodyA.pdf}
     \caption{Diagrams contributing to the process of $\pi^{-} p \rightarrow D^{\ast-} \Lambda_{c}$. Diagrams (a), (b) and (c) correspond to the $s$, $u$ and $t$-channels contributions, respectively. Here $\Lambda_c $ refers to $\Lambda_c(2286)$. \label{Fig.1}}
\end{figure*}
\begin{figure*}[htb]
     \includegraphics[width=165mm]{TwobodyB.pdf}
     \caption{Diagrams contributing to the process of $\pi^{-} p \rightarrow D^{-} \Lambda_{c} / \Lambda_{c}(2910) / \Lambda_{c}(2940)$. Diagrams (a), (b) and (c) correspond to the $s$, $u$ and $t$-channels contributions, respectively. Here $\Lambda_c $ and $\Lambda_c^\ast$ refer to $\Lambda_c(2286)$ and $\Lambda_c(2910)/\Lambda_c(2940)$, respectively. \label{Fig.2}}
\end{figure*}

As discussed in the introduction, the experimental measurement for the cross sections for $\pi^{-} p \rightarrow D^{\ast-} \Lambda_{c}$ provide us reference process to determine the model parameter involved in the present estimations. The diagrams contributing to $\pi^{-} p \rightarrow D^{\ast-} \Lambda_{c}$ are listed in Fig.~\ref{Fig.1}, where the $s-$channel nucleon exchange (diagrams (a)), $u-$channel $\Sigma^{++}_{c}$ exchange (diagrams (b)) and $t-$channel $D, D^{\ast}$ meson exchange (diagrams (c)) are considered, respectively. 

In the present calculation, we employ the effective Lagrangian approach to depict the relevant hadron interaction vertices. The effective Lagrangians for $\pi N N$, $D^{\ast} D \pi$, $D^{\ast} D^{\ast} \pi$, $D N \Sigma_{c}$, $D^{\ast} N \Sigma_{c}$, $\Lambda_{c} N D$, $\Lambda_{c} \pi \Sigma_{c}$, $\Lambda_{c} N D^{\ast}$, $\Lambda^{\ast}_{c1} N D$, $\Lambda^{\ast}_{c1} \pi \Sigma_{c}$, $\Lambda^{\ast}_{c2} N D$, $\Lambda^{\ast}_{c2} \pi \Sigma_{c}$, can be written as~\cite{Dong:2010xv, Dong:2011ys, He:2011jp, Dong:2014ksa, Xie:2015zga, Kim:2015ita, Chen:2011xk, Liu:2001ce, Okubo:1975sc, Jackson:2015dva, Oh:2000qr},
\begin{eqnarray}
\mathcal{L}_{\pi N N}&=&-i g_{\pi NN} \bar{N} \gamma_{5} \vec{\tau} \cdot \vec{\pi} N,\nonumber\\
\mathcal{L}_{D^{\ast} D \pi}&=&g_{D^{\ast} D \pi} D^{\ast}_{\mu} \bar{\tau} \cdot \Big(D \partial^{\mu} \vec{\pi} - \partial^{\mu} D \vec{\pi}\Big),\nonumber\\
\mathcal{L}_{D^{\ast} D^{\ast} \pi}&=&-g_{D^{\ast} D^{\ast} \pi} \varepsilon^{\mu \nu \alpha \beta} \partial_{\mu} D^{\ast}_{\nu} \pi \partial_{\alpha} \bar{D}^{\ast}_{\beta},\nonumber\\
\mathcal{L}_{D N \Sigma_{c}}&=&-i g_{ D N \Sigma_{c}} \bar{N} \gamma_{5} D \Sigma_{c}+H.c.,\nonumber\\ 
\mathcal{L}_{D^{\ast} N \Sigma_{c}}&=&g_{ D^{\ast} N \Sigma_{c}} \bar{N} \gamma_{\mu} D^{\ast \mu} \Sigma_{c}+H.c.,\nonumber\\ 
\mathcal{L}_{Y_{{1/2^{\pm}}} N D}&=&i g_{Y N D} \bar{Y} \left\{
\begin{array}{c}
	\gamma_5 \\
	1
\end{array}	
\right\} N D +H.c.,\nonumber\\ 
\mathcal{L}_{Y_{{1/2^{\pm}}} N D^{\ast}}&=&g_{Y N D^{\ast}} \bar{Y} \gamma^{\mu} \left\{
\begin{array}{c}
	1 \\
	\gamma_5
\end{array}	
\right\} N D^{\ast}_{\mu} +H.c.,\nonumber\\
\mathcal{L}_{Y_{{1/2^{\pm}}} \pi \Sigma_{c}}&=&i g_{Y \pi \Sigma_{c}} \bar{Y} \left\{
\begin{array}{c}
	\gamma_5 \\
	1
\end{array}	
\right\} \vec{\pi} \cdot \vec{\Sigma}_{c}+H.c.,\nonumber\\ 
\mathcal{L}_{Y_{{3/2^{\pm}}} N D}&=&\frac{g_{Y N D}}{m_{\pi}} \bar{N} \left\{
\begin{array}{c}
	\gamma_5 \\
	1
\end{array}	
\right\} Y^{\mu}  \partial^{\mu} D+H.c.,\nonumber\\
\mathcal{L}_{Y_{{3/2^{\pm}}} \pi \Sigma_{c}}&=& \frac{g_{Y \pi \Sigma_{c}}}{m_{\pi}} \bar{\Sigma}_{c} \left\{
\begin{array}{c}
	\gamma_5 \\
	1
\end{array}	
\right\} Y^{\mu} \partial^{\mu} \pi+H.c., \label{Eq.2}
\end{eqnarray}
with the upper and lower symbols in the curly bracket to be the positive and negative parity of $\Lambda_{c}$ states, respectively. The symbol $Y_{J^{p}}$ refers to the $\Lambda_{c}(2286)/\Lambda_c(2910)/\Lambda_c(2940)$ states with spin $J$ and parity $P$.

For the vertexes $\Lambda_{c}(2910) N D^{\ast}$ and $\Lambda_{c}(2940) N D^{\ast}$, we consider that the $\Lambda_{c}(2910)$ and $\Lambda_{c}(2940)$ are both $N D^{\ast}$ molecular states, and the effective Lagrangians of the molecular states coupling to their components are~\cite{Yue:2024paz},
\begin{eqnarray}
\mathcal{L}_{\Lambda_{c1}^\ast N D^{\ast}}&=& g_{\Lambda_{c1}^\ast N D^{\ast}} \bar{\Lambda}_{c1}^\ast \gamma^{\mu} \gamma_{5} D^{\ast}_{\mu} N +H.c.,\nonumber\\ 
\mathcal{L}_{\Lambda_{c2}^\ast N D^{\ast}}&=& g_{\Lambda_{c2}^{\ast} N D^{\ast}} \bar{\Lambda}_{c2}^{\ast \mu} D^{\ast}_{\mu} N +H.c.. \label{Eq.3}
\end{eqnarray}

Hereafter $\Lambda_{c1}^\ast$ and $\Lambda_{c2}^\ast$ refer to $\Lambda_c(2910)$ and $\Lambda_c(2940)$, respectively. With the above effective Lagrangians, one can obtain the amplitudes of the $\pi^{-} p \rightarrow D^{\ast-} \Lambda_{c}$ process, which are,
\begin{eqnarray}
\mathcal{M}^{s}_{n}&=&\bar{u}(p_{4}) (g_{\Lambda_{c} N D^{\ast}} \gamma^{\mu}) \mathcal{S}^{1/2}(k_{1}, m_{n}, \Gamma_{n}) (-i g_{\pi N N} \gamma_{5}) \nonumber\\ &\times& u(p_{2}) \epsilon^{\mu}(p_{3}) \Big[F(k_{1}, m_{n}, \Lambda_{r}) \Big]^2,\nonumber\\
\mathcal{M}^{u}_{\Sigma_{c}}&=&\bar{u}(p_{4}) (i g_{\Lambda_{c} \pi \Sigma_{c}} \gamma_{5}) \mathcal{S}^{1/2}(k_{1}, m_{\Sigma_{c}}, \Gamma_{\Sigma_{c}}) (g_{D^{\ast} N \Sigma_{c}} \gamma_{\mu}) \nonumber\\ &\times& u(p_{2}) \epsilon^{\mu}(p_{3}) \Big[F(k_{1}, m_{\Sigma_{c}}, \Lambda_{r}) \Big]^2,\nonumber\\
\mathcal{M}^{t}_{D}&=&\bar{u}(p_{4}) (i g_{\Lambda_{c} N D} \gamma_{5}) u(p_{2}) \mathcal{S}^{0}(k_{1}, m_{D}, \Gamma_{D}) \epsilon^{\mu}(p_{3}) \nonumber\\ &\times& \Big[g_{D^{\ast} D \pi} (-i p^{\mu}_{1}) \Big] \Big[F(k_{1}, m_{D}, \Lambda_{r}) \Big]^2,\nonumber\\
\mathcal{M}^{t}_{D^{\ast}}&=&\bar{u}(p_{4}) (g_{\Lambda_{c} N D^{\ast}} \gamma_{\nu}) u(p_{2}) \mathcal{S}^{1}_{\nu \alpha}(k_{1}, m_{D^{\ast}}, \Gamma_{D^{\ast}}) \epsilon^{\mu}(p_{3}) \nonumber\\ &\times& \Big[-g_{D^{\ast} D^{\ast} \pi} \varepsilon^{\mu \theta \alpha \beta} (p^{\theta}_{3} k^{\beta}_{1}) \Big] \Big[F(k_{1}, m_{D^{\ast}}, \Lambda_{r}) \Big]^2, \label{Eq.4}
\end{eqnarray}
where the superscripts $s$, $u$ and $t$ correspond to $s$, $u$ and $t$-channel, respectively, and the subscripts $n$, $\Sigma_{c}$, $D$ and $D^{\ast}$ correspond to the propagators in the processes.

With the observation of $\Lambda_{c}(2910)$ and $\Lambda_{c}(2940)$, we can investigate their production properties. The diagrams contributing to $\pi^{-} p \rightarrow D^{-} \Lambda_{c} / \Lambda_{c}(2910) / \Lambda_{c}(2940)$ are listed in Fig.~\ref{Fig.2}. Similarly, we also consider the relevant $s$, $u$, $t$-channel, respectively. It should be noted that only the $D^\ast$ meson exchange is valid in the $t-$channel. With the effective Lagrangians in Eq.~\eqref{Eq.2} and Eq.~\eqref{Eq.3}, one can obtain the amplitudes of the $\pi^{-} p \rightarrow D^{-} \Lambda_{c} / \Lambda_{c}(2910) / \Lambda_{c}(2940)$ process, which are,
\begin{eqnarray}
\mathcal{M}^{s}_{\Lambda_{c}}&=&\bar{u}(p_{4}) (i g_{\Lambda_{c} N D} \gamma_{5}) \mathcal{S}^{1/2}(k_{1}, m_{n}, \Gamma_{n}) (-ig_{\pi N N} \gamma_{5}) \nonumber\\ &\times& u(p_{2}) \Big[F(k_{1}, m_{n}, \Lambda_{r}) \Big]^2,\nonumber\\
\mathcal{M}^{s}_{\Lambda^{\ast}_{c1}}&=&\bar{u}(p_{4}) (g_{\Lambda^{\ast}_{c1} N D}) \mathcal{S}^{1/2}(k_{1}, m_{n}, \Gamma_{n}) (-ig_{\pi N N} \gamma_{5}) \nonumber\\ &\times& u(p_{2}) \Big[F(k_{1}, m_{n}, \Lambda_{r}) \Big]^2,\nonumber\\
\mathcal{M}^{s}_{\Lambda^{\ast}_{c2}}&=&\bar{u}^{\mu}(p_{4}) \Big(\frac{g_{\Lambda^{\ast}_{c2} N D}}{m_{\pi}} \gamma_{5} (-i p^{\mu}_{3}) \Big) \mathcal{S}^{1/2}(k_{1}, m_{n}, \Gamma_{n}) \nonumber\\ &\times& (-ig_{\pi N N} \gamma_{5}) u(p_{2}) \Big[F(k_{1}, m_{n}, \Lambda_{r}) \Big]^2,\nonumber\\
\mathcal{M}^{u}_{\Lambda_{c}}&=&\bar{u}(p_{4}) (i g_{\Lambda_{c} \pi \Sigma_{c}} \gamma_{5}) \mathcal{S}^{1/2}(k_{1}, m_{\Sigma_{c}}, \Gamma_{\Sigma_{c}}) (-ig_{D N \Sigma_{c}} \gamma_{5}) \nonumber\\ &\times& u(p_{2}) \Big[F(k_{1}, m_{\Sigma_{c}}, \Lambda_{r}) \Big]^2,\nonumber\\
\mathcal{M}^{u}_{\Lambda^{\ast}_{c1}}&=&\bar{u}(p_{4}) (i g_{\Lambda^{\ast}_{c1} \pi \Sigma_{c}}) \mathcal{S}^{1/2}(k_{1}, m_{\Sigma_{c}}, \Gamma_{\Sigma_{c}}) (-ig_{D N \Sigma_{c}} \gamma_{5}) \nonumber\\ &\times& u(p_{2}) \Big[F(k_{1}, m_{\Sigma_{c}}, \Lambda_{r}) \Big]^2,\nonumber\\
\mathcal{M}^{u}_{\Lambda^{\ast}_{c2}}&=&\bar{u}^{\mu}(p_{4}) \Big(\frac{g_{\Lambda^{\ast}_{c2} \pi \Sigma_{c}}}{m_{\pi}} \gamma_{5} (-i p^{\mu}_{1}) \Big) \mathcal{S}^{1/2}(k_{1}, m_{\Sigma_{c}}, \Gamma_{\Sigma_{c}}) \nonumber\\ &\times& (-ig_{D N \Sigma_{c}} \gamma_{5}) u(p_{2}) \Big[F(k_{1}, m_{\Sigma_{c}}, \Lambda_{r}) \Big]^2,\nonumber\\
\mathcal{M}^{t}_{\Lambda_{c}}&=&\bar{u}(p_{4}) (g_{\Lambda_{c} N D^{\ast}} \gamma_{\mu}) u(p_{2}) \mathcal{S}^{1}_{\mu \nu}(k_{1}, m_{D^{\ast}}, \Gamma_{D^{\ast}}) \nonumber\\ &\times& \Big( g_{D^{\ast} D \pi} (-i p^{\nu}_{1}) \Big) \Big[F(k_{1}, m_{D^{\ast}}, \Lambda_{r}) \Big]^2,\nonumber\\
\mathcal{M}^{t}_{\Lambda^{\ast}_{c1}}&=&\bar{u}(p_{4}) (g_{\Lambda^{\ast}_{c1} N D^{\ast}} \gamma_{5} \gamma_{\mu}) u(p_{2}) \mathcal{S}^{1}_{\mu \nu}(k_{1}, m_{D^{\ast}}, \Gamma_{D^{\ast}}) \nonumber\\ &\times& \Big( g_{D^{\ast} D \pi} (-i p^{\nu}_{1}) \Big) \Big[F(k_{1}, m_{D^{\ast}}, \Lambda_{r}) \Big]^2,\nonumber
\end{eqnarray}
\begin{eqnarray}
\mathcal{M}^{t}_{\Lambda^{\ast}_{c2}}&=&\bar{u}^{\mu}(p_{4}) (g_{\Lambda^{\ast}_{c2} N D^{\ast}} ) u(p_{2}) \mathcal{S}^{1}_{\mu \nu}(k_{1}, m_{D^{\ast}}, \Gamma_{D^{\ast}}) \nonumber\\ &\times& \Big( g_{D^{\ast} D \pi} (-i p^{\nu}_{1}) \Big) \Big[F(k_{1}, m_{D^{\ast}}, \Lambda_{r}) \Big]^2, \label{Eq.5}
\end{eqnarray}
where the superscripts $s$, $u$ and $t$ correspond to $s$, $u$ and $t$-channel, respectively, whlie the subscripts $\Lambda_c $, $\Lambda^{\ast}_{c1}$ $\Lambda^{\ast}_{c2}$ refer to the final states of the process, respectively.

In the above amplitudes, $\mathcal{S}^{0}(k_i,m_i,\Gamma_i)$ and $\mathcal{S}^{1}_{\mu \nu}(k_{i}, m_{i}, \Gamma_{i})$ are the propagators of scalar and vector meson with four momentum $k_i$, mass $m_i$ and width $\Gamma_i$, respectively, and the concrete expressions are, 
\begin{eqnarray}
\mathcal{S}^{0}(k_i,m_i,\Gamma_i)&=&\frac{i} {k^2_{i}-m^2_{i} +i m_i \Gamma_{i}},\nonumber \\
\mathcal{S}^{1}_{\mu \nu}(k_{i}, m_{i}, \Gamma_{i}) &=& \frac{-g^{\mu \nu}+(k^{\mu}_{i} k^{\nu}_{i} / m^{2}_{i})}{k^{2}_{i}-m^{2}_{i}+i m_{i} \Gamma_{i}}, \label{Eq.6}
\end{eqnarray}

Then, $\mathcal{S}^{1/2}(k_{i},m_{i},\Gamma_{i})$ is the propagator of $n$, $\Lambda_{c}(2286)$, $\Lambda_{c}(2910)$ and $\Sigma^{++}_{c}$, whose spin of $1/2$. While $\mathcal{S}^{3/2}_{\mu \nu}(k_{i},m_{i},\Gamma_{i})$ is the propagator of $\Lambda_{c}(2940)$.  The detailed expressions of $\mathcal{S}^{1/2}(k_{i},m_{i},\Gamma_{i})$ and $\mathcal{S}^{3/2}_{\mu \nu}(k_{i},m_{i},\Gamma_{i})$ read,
\begin{eqnarray}
\mathcal{S}^{1/2}(k_{i},m_{i},\Gamma_{i}) = \frac{\slash\!\!\!k_{i}+m_{i}} {k^2_{i}-m^2_{i} +i m_{i} \Gamma_{i}},\nonumber
\end{eqnarray}
\begin{eqnarray}
&&\mathcal{S}^{3/2}_{\mu \nu}(k_{i},m_{i},\Gamma_{i})  = \frac{\slash\!\!\!k_{i}+m_{i}} {k^2_{i}-m^2_{i} +i m_{i} \Gamma_{i}} \nonumber\\ &&\qquad \qquad\times  \Big(-g^{\mu \nu} + \frac{\gamma^{\mu} \gamma^{\nu}}{3} + \frac{2 k^{\mu} k^{\nu}}{3 m^{2}_{i}} +\frac{\gamma^{\mu} k^{\nu}-k^{\mu} \gamma^{\nu}}{3 m_{i}} \Big).\qquad \label{Eq.7}
\end{eqnarray}

In addition, the form factor $F(k_{i},m_{i},\Lambda_{r})$ is introduced to depict the inner structure of the involved hadrons in each vertex, its specific expression is,
\begin{eqnarray}
F(k_{i},m_{i},\Lambda_{r})
=\frac{\Lambda^{4}_{r}} {\Lambda^{4} _{r} +(k^{2}_{i}-m^{2}_{i})^2},\label{Eq.8},
\end{eqnarray}
where $k_{i}$ and $m_{i}$ are the four momentum and the mass of the exchanged hadron, respectively. $\Lambda_{r}$ is a phenomeno-logical model parameter and the specific value will be discussed in Section \ref{sec:MA}.

\begin{figure*}[htb]
     \includegraphics[width=165mm]{Threebody.pdf}
	\caption{Diagrams contributing to the process of $\pi^{-} p \rightarrow D^{-} D^{0} p$. Diagrams (a), (b) and (c) correspond to the $s$, $u$ and $t$-channels contributions, respectively. Here $\Lambda_c $ and $\Lambda_c^\ast$ refer to $\Lambda_c(2286)$ and $\Lambda_c(2910)/\Lambda_c(2940)$, respectively. \label{Fig.3}}
\end{figure*}

\subsection{Cross Sections for $\pi^{-} p \rightarrow D^{-} D^{0} p$}
Since $\Lambda_c(2910)$ and $\Lambda_c(2940)$ can decay into $D^0 p$ final states, then we can investigate $\Lambda_c(2910)$ and $\Lambda_c(2940)$ in the $\pi^{-} p \rightarrow D^{-} D^{0} p$ process. The diagrams contributing to $\pi^{-} p \rightarrow D^{-} D^{0} p$ are listed in Fig.~\ref{Fig.3}. It should be noted that $\Lambda_c(2286)$ could play the role of background since $\Lambda_c(2286)$ is below the threshold of $D^0 p$ in this process. With the above effective Lagrangians in Eq.~\eqref{Eq.2} and Eq.~\eqref{Eq.3}, one can also obtain the amplitudes corresponding to the $\pi^{-} p \rightarrow D^{-} D^{0} p$ processes, which are,
\begin{widetext}
	\begin{eqnarray}
\mathcal{M^{\prime}}^{s}_{\Lambda_{c}}&=&\bar{u}(p_{5}) (i g_{\Lambda_{c} N D} \gamma_{5}) \mathcal{S}^{1/2}(k_{2}, m_{\Lambda_{c}}, \Gamma_{\Lambda_{c}}) (i g_{\Lambda_{c} N D} \gamma_{5}) \mathcal{S}^{1/2}(k_{1}, m_{n}, \Gamma_{n}) (-i g_{\pi N N} \gamma_{5}) u(p_{2}) \nonumber\\ &\times& F(k_{2}, m_{\Lambda_{c}}, \Lambda_{r}) \Big[F(k_{1}, m_{n}, \Lambda_{r}) \Big]^2,\nonumber\\
\mathcal{M^{\prime}}^{s}_{\Lambda^{\ast}_{c1}}&=&\bar{u}(p_{5}) (g_{\Lambda^{\ast}_{c1} N D}) \mathcal{S}^{1/2}(k_{2}, m_{\Lambda^{\ast}_{c1}}, \Gamma_{\Lambda^{\ast}_{c1}}) (g_{\Lambda^{\ast}_{c1} N D}) \mathcal{S}^{1/2}(k_{1}, m_{n}, \Gamma_{n}) (-i g_{\pi N N} \gamma_{5}) u(p_{2}) \nonumber\\ &\times& F(k_{2}, m_{\Lambda^{\ast}_{c1}}, \Lambda_{r}) \Big[F(k_{1}, m_{n}, \Lambda_{r}) \Big]^2,\nonumber\\
\mathcal{M^{\prime}}^{s}_{\Lambda^{\ast}_{c2}}&=&\bar{u}(p_{5}) \Big[\frac{ g_{\Lambda^{\ast}_{c2} N D}}{m_{\pi}} \gamma_{5} (-i p^{\mu}_{4})\Big] \mathcal{S}^{3/2}_{\mu \nu}(k_{2}, m_{\Lambda^{\ast}_{c2}}, \Gamma_{\Lambda^{\ast}_{c2}}) \Big[\frac{ g_{\Lambda^{\ast}_{c2} N D}}{m_{\pi}} \gamma_{5} (-i p^{\nu}_{3})\Big] \mathcal{S}^{1/2}(k_{1}, m_{n}, \Gamma_{n}) (-i g_{\pi N N} \gamma_{5}) u(p_{2}) \nonumber\\ &\times& F(k_{2}, m_{\Lambda^{\ast}_{c2}}, \Lambda_{r}) \Big[F(k_{1}, m_{n}, \Lambda_{r}) \Big]^2,\nonumber\\
\mathcal{M^{\prime}}^{u}_{\Lambda_{c}}&=&\bar{u}(p_{5}) (i g_{\Lambda_{c} N D} \gamma_{5}) \mathcal{S}^{1/2}(k_{2}, m_{\Lambda_{c}}, \Gamma_{\Lambda_{c}}) (i g_{\Lambda_{c} \pi \Sigma_{c}} \gamma_{5}) \mathcal{S}^{1/2}(k_{1}, m_{\Sigma_{c}}, \Gamma_{\Sigma_{c}}) (-i g_{D N \Sigma_{c}} \gamma_{5}) u(p_{2}) \nonumber\\ &\times& F(k_{2}, m_{\Lambda_{c}}, \Lambda_{r}) \Big[F(k_{1}, m_{\Sigma_{c}}, \Lambda_{r}) \Big]^2,\nonumber\\
\mathcal{M^{\prime}}^{u}_{\Lambda^{\ast}_{c1}}&=&\bar{u}(p_{5}) (g_{\Lambda^{\ast}_{c1} N D}) \mathcal{S}^{1/2}(k_{2}, m_{\Lambda^{\ast}_{c1}}, \Gamma_{\Lambda^{\ast}_{c1}}) (i g_{\Lambda^{\ast}_{c1} \pi \Sigma_{c}}) \mathcal{S}^{1/2}(k_{1}, m_{\Sigma_{c}}, \Gamma_{\Sigma_{c}}) (-i g_{D N \Sigma_{c}} \gamma_{5}) u(p_{2}) \nonumber\\ &\times& F(k_{2}, m_{\Lambda^{\ast}_{c1}}, \Lambda_{r}) \Big[F(k_{1}, m_{\Sigma_{c}}, \Lambda_{r}) \Big]^2,\nonumber\\
\mathcal{M^{\prime}}^{u}_{\Lambda^{\ast}_{c2}}&=&\bar{u}(p_{5}) \Big[\frac{ g_{\Lambda^{\ast}_{c2} N D}}{m_{\pi}} \gamma_{5} (-i p^{\mu}_{4})\Big] \mathcal{S}^{3/2}_{\mu \nu}(k_{2}, m_{\Lambda^{\ast}_{c2}}, \Gamma_{\Lambda^{\ast}_{c2}}) \Big[\frac{ g_{\Lambda^{\ast}_{c2} \pi \Sigma_{c}}}{m_{\pi}} \gamma_{5} (-i p^{\nu}_{1})\Big] \mathcal{S}^{1/2}(k_{1}, m_{\Sigma_{c}}, \Gamma_{\Sigma_{c}}) (-i g_{D N \Sigma_{c}} \gamma_{5}) u(p_{2}) \nonumber\\ &\times& F(k_{2}, m_{\Lambda^{\ast}_{c2}}, \Lambda_{r}) \Big[F(k_{1}, m_{\Sigma_{c}}, \Lambda_{r}) \Big]^2,\nonumber
\end{eqnarray}
\begin{eqnarray}
\mathcal{M^{\prime}}^{t}_{\Lambda_{c}}&=&\bar{u}(p_{5}) (i g_{\Lambda_{c} N D} \gamma_{5}) \mathcal{S}^{1/2}(k_{2}, m_{\Lambda_{c}}, \Gamma_{\Lambda_{c}}) (g_{\Lambda_{c} N D^{\ast}} \gamma^{\mu}) \mathcal{S}^{1}_{\mu \nu}(k_{1}, m_{D^{\ast}}, \Gamma_{D^{\ast}}) [g_{D^{\ast} D \pi}(-i p^{\nu}_{1})] u(p_{2}) \nonumber\\ &\times& F(k_{2}, m_{\Lambda_{c}}, \Lambda_{r}) \Big[F(k_{1}, m_{D^{\ast}}, \Lambda_{r}) \Big]^2,\nonumber\\
\mathcal{M^{\prime}}^{t}_{\Lambda^{\ast}_{c1}}&=&\bar{u}(p_{5}) (g_{\Lambda^{\ast}_{c1} N D}) \mathcal{S}^{1/2}(k_{2}, m_{\Lambda^{\ast}_{c1}}, \Gamma_{\Lambda^{\ast}_{c1}}) (g_{\Lambda^{\ast}_{c1} N D^{\ast} \gamma_{5} \gamma^{\mu}}) \mathcal{S}^{1}_{\mu \nu}(k_{1}, m_{D^{\ast}}, \Gamma_{D^{\ast}}) [g_{D^{\ast} D \pi}(-i p^{\nu}_{1})] u(p_{2}) \nonumber\\ &\times& F(k_{2}, m_{\Lambda^{\ast}_{c1}}, \Lambda_{r}) \Big[F(k_{1}, m_{D^{\ast}}, \Lambda_{r}) \Big]^2,\nonumber\\
\mathcal{M^{\prime}}^{t}_{\Lambda^{\ast}_{c2}}&=&\bar{u}(p_{5}) \Big[\frac{ g_{\Lambda^{\ast}_{c2} N D}}{m_{\pi}} \gamma_{5} (-i p^{\mu}_{4})\Big] \mathcal{S}^{3/2}_{\mu \nu}(k_{2}, m_{\Lambda^{\ast}_{c2}}, \Gamma_{\Lambda^{\ast}_{c2}}) (g_{\Lambda^{\ast}_{c2} N D^{\ast}}) \mathcal{S}^{1}_{\nu \rho}(k_{1}, m_{D^{\ast}}, \Gamma_{D^{\ast}}) [g_{D^{\ast} D \pi} (-i p^{\rho}_{1})] u(p_{2}) \nonumber\\ &\times& F(k_{2}, m_{\Lambda^{\ast}_{c2}}, \Lambda_{r}) \Big[F(k_{1}, m_{D^{\ast}}, \Lambda_{r}) \Big]^2, \label{Eq.10}
\end{eqnarray}

\end{widetext}
where the superscripts $s$, $u$ and $t$ correspond to $s$, $u$ and $t$-channel, respectively. In the above amplitudes, the form of propagators and form factor are consistent with those in Eqs.~\eqref{Eq.6}-\eqref{Eq.8}.

\section{NUMERICAL RESULTS AND DISCUSSIONS}
\label{sec:MA}
\subsection{Coupling Constants}
Before the estimations of the cross sections, the values of coupling constants should be clarified. The coupling constants $g_{\pi N N}$, $g_{\Lambda_{c} N D}$, $g_{\Lambda_{c} \pi \Sigma_{c}}$, $g_{D N \Sigma_{c}}$, $g_{D^{\ast} N \Sigma_{c}}$ and $g_{\Lambda_{c} N D^{\ast}}$ can be determined by SU(4) symmetry~\cite{Xie:2015zga, Dong:2010xv, Dong:2011ys, He:2011jp, Dong:2014ksa, Chen:2011xk, Liu:2001ce, Okubo:1975sc}, and the specific values are collected in Table~\ref{Tab:1}. It should be noted that the large mass differences between the charmed quark and the light quarks significantly break SU(4) flavor symmetry. Despite this limitation, theorists often invoke this symmetry to estimate coupling constants, particularly those involving charmed baryons, due to the lack of alternative systematic methods. The coupling constant $g_{D^{\ast} D^{\ast} \pi}$ can be determined by the heavy quark spin symmetry and the value is 9.08~\cite{Oh:2000qr}. In addition, the coupling constants $g_{\Lambda^{\ast}_{c1} N D}$, $g_{\Lambda^{\ast}_{c1} \pi \Sigma_{c}}$, $g_{\Lambda^{\ast}_{c2} N D}$, $g_{\Lambda^{\ast}_{c2} \pi \Sigma_{c}}$ and $g_{D^{\ast} D \pi}$ can be determined by combining the related effective Lagrangians and the corresponding decay widths of the corresponding decay processes. Based on the effective Lagrangians, one can obtain the corresponding amplitude $\mathcal{M}_{A\to BC}$. Then, the decay width of the decay process $A\to BC$ can be written as,
\begin{eqnarray}
\Gamma_{A \to BC} = \frac{1}{(2J+1)8\pi} \frac{|\vec{k}_f|}{M^{2}} \overline{|\mathcal{M}_{A\to BC}|^2} \label{Eq:2BDecay}
\end{eqnarray}
where $M$ and $J$ are the mass and angular momentum of the initial states. $\vec{k}_f$ is the three momentum of the final states in the initial rest frame. In addition, in Ref.~\cite{Yue:2024paz}, the decay properties of $\Lambda_c(2910)$ and $\Lambda_c(2940)$ were investigated, and the estimations indicated that the branching fractions of $N D$ and $\pi \Sigma_{c}$ channels for $\Lambda_{c}(2910)$ and $\Lambda_{c}(2940)$ are about,
\begin{eqnarray}
&&\mathrm{Br}(\Lambda_{c}(2910) \rightarrow N D)=40 \%,\nonumber\\ 
&& \mathrm{Br}(\Lambda_{c}(2910) \rightarrow \pi \Sigma)=60 \%, \nonumber\\ 
&&\mathrm{Br}(\Lambda_{c}((2940)) \rightarrow N D)=11 \%, \nonumber \\
&&\mathrm{Br}(\Lambda_{c}(2940) \rightarrow \pi \Sigma)=12.5 \%. 
\end{eqnarray}

With the above branching fractions, the central values of the widths of $\Lambda_c(2940)$ and $\Lambda_c(2910)$ and the formula in Eq.~\eqref{Eq:2BDecay}, one can obtain the coupling constants $g_{\Lambda_{c1}^\ast ND}$, $g_{\Lambda_{c1}^\ast \Sigma_c \pi }$, $g_{\Lambda_{c2}^\ast ND}$, and $g_{\Lambda_{c2}^\ast \Sigma_c \pi }$, which are listed in Table~\ref{Tab:1}. As for the coupling constant $g_{D^{\ast} D \pi}$, we take the same value as the one in Ref.~\cite{Xie:2015zga}, which is $g_{D^\ast D\pi }=14.1$.

In addition to the above coupling constants, the coupling constants relevant to $\Lambda_c(2910)/\Lambda_c(2940)$ and their components $D^\ast N$  could be determined by the compositeness condition with non-relativistic limit~\cite{Weinberg:1965zz, Baru:2003qq}, which is
\begin{eqnarray}
g_{\Lambda^{\ast}_{c} N D^{\ast}}^{2}
=\frac{4 \pi} {4 M_{\Lambda^{\ast}_{c}} m_{N}} \frac{(m_{D^{\ast}} + m_{N})^{5/2}} {(m_{D^{\ast}} m_{N})^{1/2}} \sqrt{32 \epsilon},
\end{eqnarray}
with $\epsilon = m_{N} + m_{D^{\ast}} - m_{\Lambda^{\ast}_{c}}$ to be the binding energy of $\Lambda_{c}(2910)/\Lambda_c(2940)$. Here, we take $\epsilon = 32 \; \mathrm{MeV}$ for $\Lambda_{c} (2910)$ and $\epsilon = 6.2 \; \mathrm{MeV}$ for $\Lambda_{c}(2940)$. The concrete values of $g_{\Lambda_{c}^\ast ND^\ast}$ are also presented in Table~\ref{Tab:1}.

\begin{table}
\caption{The coupling constants involved in the $ \pi^{-} p \rightarrow D^{-} D^{0} p$ process.}
\label{Tab:1}
\renewcommand{\arraystretch}{2}
\setlength{\tabcolsep}{5pt}
\centering
\begin{tabular}{cccc}
\toprule[1pt]
Coupling constant&Value& Coupling constant&Value\\
\midrule[1pt]
$g_{\pi NN}$&$13.45$&
$g_{\Lambda^{\ast}_{c1} N D }$&$0.99$\\
$g_{\Lambda_{c} N D}$&$-13.98$&
$g_{\Lambda^{\ast}_{c2} N D}$&$0.84$\\
$g_{\Lambda_{c} \pi \Sigma_{c}}$&$9.32$&
$g_{\Lambda^{\ast}_{c1}  \pi \Sigma_{c}}$&$2.22$\\
$g_{D N \Sigma_{c}}$&$-2.69$&
$g_{\Lambda^{\ast}_{c2} \pi \Sigma_{c}}$&$1.35$\\
$g_{D^{\ast} N \Sigma_{c}}$&$3.75$&
$g_{D^{\ast} D^{\ast} \pi}$&$9.08$\\
$g_{\Lambda_{c} N D^{\ast}}$&$-5.20$&
$g_{\Lambda^{\ast}_{c1} N D^{\ast} }$&$3.55$\\
$g_{D^{\ast} D \pi}$&$14.1$&
$g_{\Lambda^{\ast}_{c2} N D^{\ast} }$&$2.34$\\
\bottomrule[1pt] 
\end{tabular} 
\end{table}

\subsection{Cross Sections for $\pi^{-} p \rightarrow D^{\ast-} \Lambda_{c}$ and $\pi^{-} p \rightarrow D^{-} \Lambda_{c} / \Lambda_{c}(2910) / \Lambda_{c}(2940)$}

\begin{figure*}[htb]
\centering
     \includegraphics[width=180mm]{TwobodycrosssectionA.pdf}
     \caption{(Color online) The cross sections for the $\pi^{-} p \rightarrow D^{\ast-} \Lambda_{c}$ process depending on the momentum of the incident pion beam. Diagram (a) shows the total cross sections for $\pi^{-} p \rightarrow D^{\ast-} \Lambda_{c}$, the black solid curve are obtained with $\Lambda_r=1.1$ GeV, while the cyan band are the uncertainties resulted from the varying of $\Lambda_r$ from 1.0 to 1.2 GeV. The red pentagram represents the upper limit of cross section $\sigma=7$ nb at $P_{\pi}=13$ GeV from Ref.~\cite{Christenson:1985ms}. Diagram (b) corresponds to the individual contributions from $s$, $u$ and $t$-channels, respectively. \label{Fig.4}}

\end{figure*}

After determining the coupling parameters, we need determine the model parameter $\Lambda_{r}$ introduced by the form factor $F(k_{i},m_{i},\Lambda_{r})$. As discussed in the above sections, $\pi^{-} p \rightarrow D^{\ast-} \Lambda_{c}$ could perform as a reference channel, since the upper limit of the cross sections have been measured at $P_{\pi}=13$ GeV. Thus, we estimate the cross sections for $\pi^{-} p \rightarrow D^{\ast-} \Lambda_{c}$ with the amplitudes in Eq.~\eqref{Eq.4}. The total amplitude of the $\pi^{-} p \rightarrow D^{\ast-} \Lambda_{c}$ can be expressed as,
\begin{eqnarray}
\mathcal{M}_{\mathrm{Tot}}&=&\mathcal{M}^{s}_{n}+\mathcal{M}^{u}_{\Sigma_{c}}+\mathcal{M}^{t}_{D}+\mathcal{M}^{t}_{D^{\ast}}
\end{eqnarray}
and with this total amplitude, the differential cross sections depending on $\cos \theta $ for $\pi^{-} p \rightarrow D^{\ast-} \Lambda_{c}$ can be obtained by, 
\begin{eqnarray}
	\frac{d{\sigma}} {d \cos\theta}
	=\frac{1} {32 \pi s} \frac{|\vec{p}_f|} {|\vec{p}_i|}  \left(\frac{1} {2} \left|\overline{\mathcal{M}_{\mathrm{Tot}}} \right|^2\right),\label{Eq:CS}
\end{eqnarray}
where $s$ and $\theta$ refer to the square of the center of mass energy and the scattering angle, which is the angle of outgoing $D^{\ast-}$ and the pion beam direction in the center direction. The $\vec{p_f}$ and $\vec{p_i}$ stand for three momenta of the final $D^{\ast-}$ meson and the initial pion beam in the center of mass system, respectively.

\begin{figure}[htb]
	\centering
     \includegraphics[width=85mm]{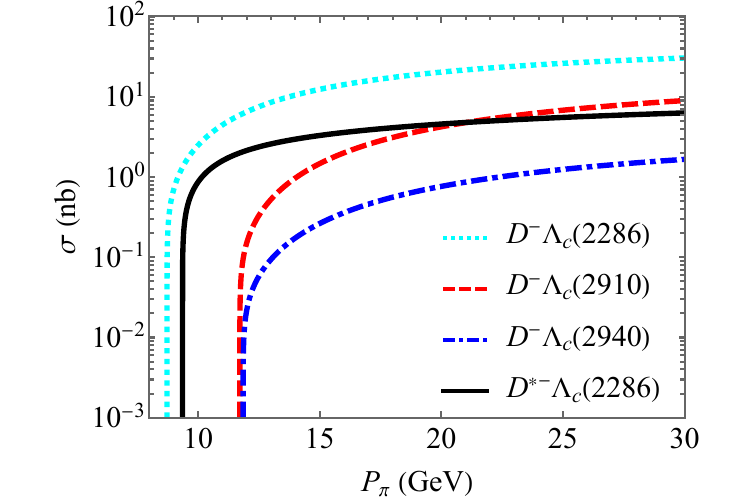}
     \caption{(Color online) The cross sections for the $\pi^{-} p \rightarrow D^{-} \Lambda_{c} / \Lambda_{c}(2910) / \Lambda_{c}(2940)$ and $\pi^{-} p \rightarrow D^{\ast-} \Lambda_{c}$ processes depending on the momentum of the incident pion beam with $\Lambda_r=1.1$ GeV. The cyan dotted, red dashed, blue dash-dotted and black solid curves correspond to the contributions from $D^{-}\Lambda_c(2286)$, $D^{-}\Lambda_c(2910)$, $D^{-}\Lambda_c(2940)$ and $D^{\ast-}\Lambda_c(2286)$ final states of the relevant processes, respectively.} 
	\label{Fig.5}

\end{figure}

\begin{figure*}[htb]
     \includegraphics[width=180mm]{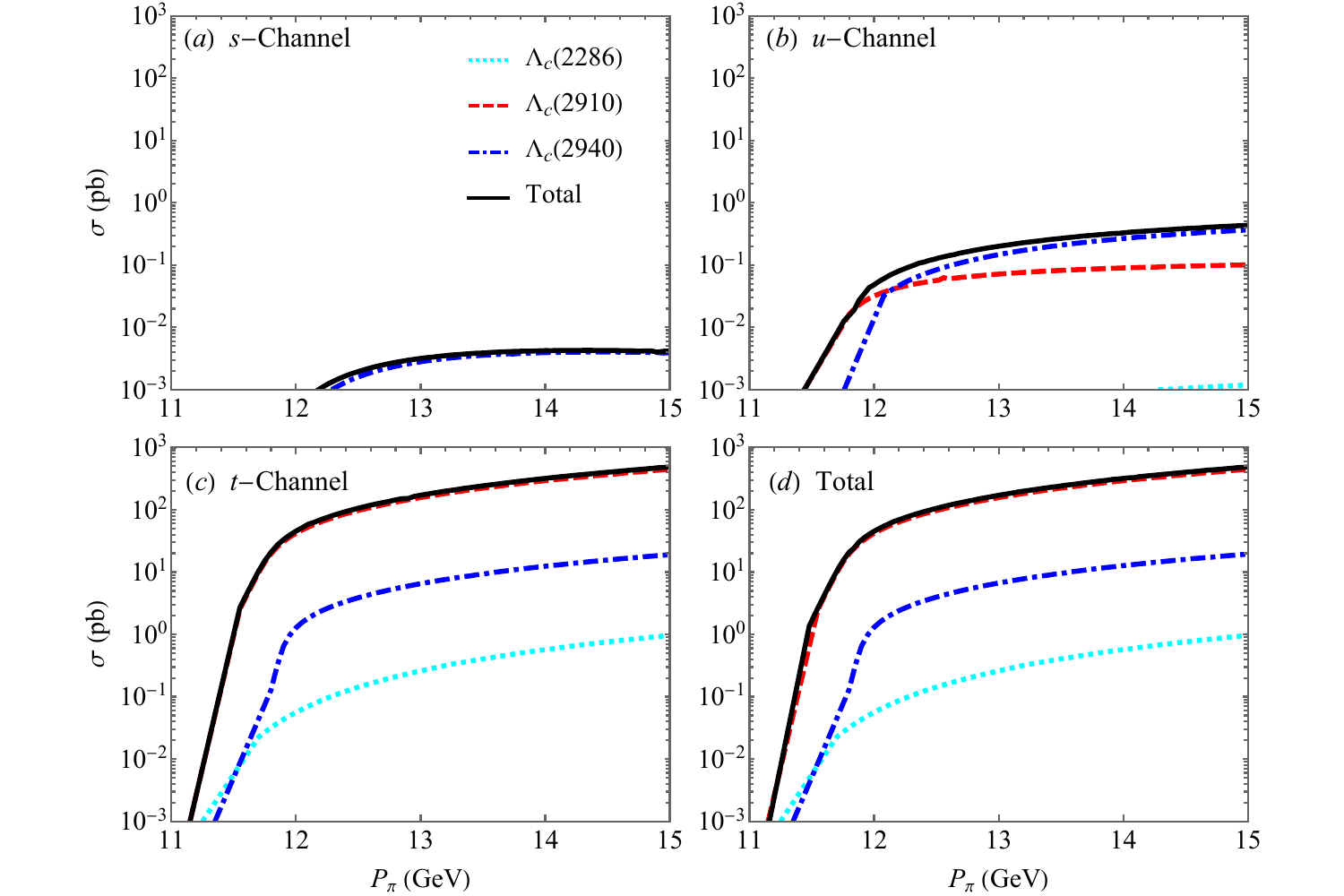}
     \caption{(Color online) The cross sections for the $\pi^{-} p \rightarrow D^{-} D^{0} p$ process depending on the momentum of the incident pion beam. Diagram (a), (b), (c) correspond to the individual contributions from $s$, $u$, and $t$-channels, respectively. While diagram (d) shows the individual contributions from $\Lambda_c(2286)$, $\Lambda_c(2910)$ and $\Lambda_c(2940)$.  \label{Fig:2}}
\end{figure*}

With the above preparations, the cross sections for $\pi^{-} p \rightarrow D^{\ast-} \Lambda_{c}$ depending on the momentum of the incident pion beam are presented in Fig.~\ref{Fig.4}. Among them, Fig.~\ref{Fig.4}(a) corresponds to the total cross sections for the $\pi^{-} p \rightarrow D^{\ast-} \Lambda_{c}$ process, the black solid curves represent the cross sections obtained with $\Lambda_r=1.1$ GeV, while the cyan band are the uncertainties resulted from the varying of $\Lambda_r$ from 1.0 to 1.2 GeV. The center position of red pentagram corresponds to the upper limit of cross section $\sigma=7$ nb at $P_{\pi}=13$ GeV~\cite{Christenson:1985ms}. In particular, the cross section is estimated to be $\Big(2.59^{+7.15}_{-2.00}\Big)$ nb at $P_{\pi}=13$ GeV in the consider parameter range. In  Fig.~\ref{Fig.4}(b), the cyan dotted, red dashed and blue dash-dotted curves stand for the individual contributions to cross sections from the $s$, $u$ and $t$-channel, respectively, where the parameter is set to be $\Lambda_{r}=1.1$ GeV.  Our results show that the contributions from $s$-channel are negligible, while the main contributions come from $t$-channel. In summary, our estimations show that the cross section for $\pi^{-} p \rightarrow D^{\ast-} \Lambda_{c}$ is $2.59$ nb at $P_{\pi}=13$ GeV with $\Lambda_r=1.1$ GeV, which is safely under the experimental upper limit reported in Ref.~\cite{Christenson:1985ms}. Thus, in the following we will adopt $\Lambda_r=1.1$ GeV to discuss other involved processes.

In addition to $\Lambda_{c}(2286)$, we also investigate the productions of $\Lambda_{c}(2910)$ and $\Lambda_{c}(2940)$ in the $\pi^- p$ scattering processes, with the amplitudes in Eq.~\eqref{Eq.5}, one can estimate the cross sections for the $\pi^{-} p \rightarrow D^{-} \Lambda_{c} / \Lambda_{c}(2910) / \Lambda_{c}(2940)$ processes, the relevant total amplitudes can be expressed as,
\begin{eqnarray}
\mathcal{M}_{\mathrm{Tot-\Lambda_{c}}}&=&\mathcal{M}^{s}_{\Lambda_{c}}+\mathcal{M}^{u}_{\Lambda_{c}}+\mathcal{M}^{t}_{\Lambda_{c}}, \nonumber\\
\mathcal{M}_{\mathrm{Tot-\Lambda^{\ast}_{c1}}}&=&\mathcal{M}^{s}_{\Lambda^{\ast}_{c1}}+\mathcal{M}^{u}_{\Lambda^{\ast}_{c1}}+\mathcal{M}^{t}_{\Lambda^{\ast}_{c1}}, \nonumber\\
\mathcal{M}_{\mathrm{Tot-\Lambda^{\ast}_{c2}}}&=&\mathcal{M}^{s}_{\Lambda^{\ast}_{c2}}+\mathcal{M}^{u}_{\Lambda^{\ast}_{c2}}+\mathcal{M}^{t}_{\Lambda^{\ast}_{c2}}, \label{Eq.16}
\end{eqnarray}

The cross sections for $\pi^{-} p \rightarrow D^{-} \Lambda_{c} / \Lambda_{c}(2910) / \Lambda_{c}(2940)$ and $\pi^{-} p \rightarrow D^{\ast-} \Lambda_{c}$ processes depending on the momentum of the incident pion beam with $\Lambda_r=1.1$ GeV are presented in Fig.~\ref{Fig.5}, where the cyan dotted, red dashed, blue dash-dotted and black solid curves correspond to the cross section for $D^{-}\Lambda_c(2286)$, $D^{-}\Lambda_c(2910)$, $D^{-}\Lambda_c(2940)$, and $D^{\ast-}\Lambda_c(2286)$ final states, respectively. From the figure, one can find that each cross section curve increases rapidly near the threshold and then becomes weakly dependent on the momentum of the incident pion beam. In addition, our results show that the cross sections for $D^{-} \Lambda_{c}(2286)$ final states are larger than that of $D^{-} \Lambda_{c}(2910)$ and $D^{-} \Lambda_{c}(2940)$, while the cross sections for $D^{\ast-} \Lambda_{c}(2286)$ final states are similar to that of $D^{-} \Lambda_{c}(2910)$. At $p_\pi=30$ GeV, the ratios of the cross sections for $\pi^- p \to D^{\ast-} \Lambda_c(2286)$, $\pi^- p \to D^{-} \Lambda_c(2286)$, $\pi^- p \to D^{-} \Lambda_c(2910)$, and $\pi^- p \to D^{-} \Lambda_c(2940)$ are estimated to be $1:4.8:1.42:0.26$.

\subsection{Cross Sections for $\pi^{-} p \rightarrow D^{-} D^{0} p$}

\begin{figure}[htb]
	\centering
     \includegraphics[width=85mm]{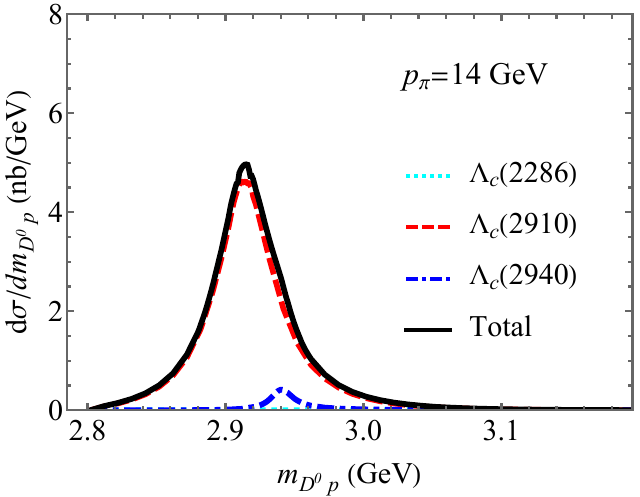}
 \caption{(Color online) The $D^0 p$ invariant mass distributions of the $\pi^{-} p \rightarrow D^{-} D^{0} p$ process at $p_{\pi}=14$ $\mathrm{GeV}$.} 
	\label{Fig:4}
\end{figure}

\begin{figure}[htb]
	\centering
	\includegraphics[width=85 mm]{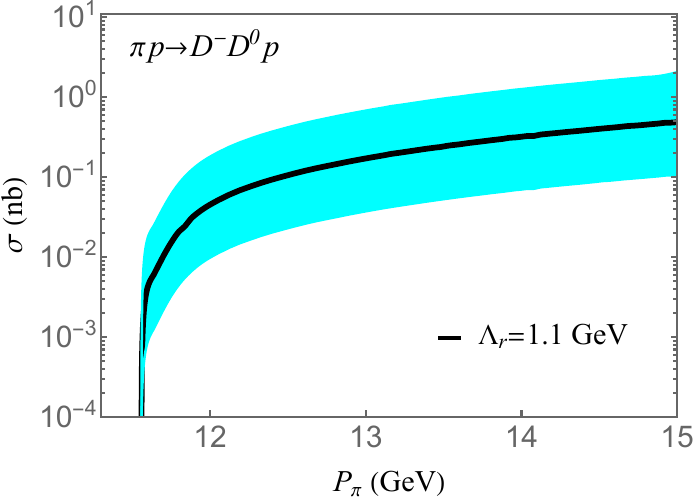}	\caption{(Color online) The total cross sections for the $\pi^{-} p \rightarrow D^{-} D^{0} p$ process depending on the momentum of the incident pion beam. The black solid curve are obtained with $\Lambda_r=1.1$ GeV, while the cyan band are the uncertainties resulted from the varying of $\Lambda_r$ from 1.0 to 1.2 GeV.} 
	\label{Fig:3}
\end{figure}

Since the $\Lambda_{c}(2910)$ and $\Lambda_{c}(2940)$ state can further decay into $D^{0}p$, so one can estimate the cross sections for the $\pi^{-} p \rightarrow D^{-} D^{0} p$ process. With the amplitudes in Eq.~\eqref{Eq.10}, the total amplitude of the $\pi^{-} p \rightarrow D^{-} D^{0} p$ process can be expressed as,
\begin{eqnarray}
\mathcal{M^{\prime}}_{\mathrm{Tot}}&=&\mathcal{M^{\prime}}^{s}_{\Lambda_{c}} + \mathcal{M^{\prime}}^{s}_{\Lambda^{\ast}_{c1}} + \mathcal{M^{\prime}}^{s}_{\Lambda^{\ast}_{c2}} + \mathcal{M^{\prime}}^{u}_{\Lambda_{c}} + \mathcal{M^{\prime}}^{u}_{\Lambda^{\ast}_{c1}} \nonumber\\ &+& \mathcal{M^{\prime}}^{u}_{\Lambda^{\ast}_{c2}} + \mathcal{M^{\prime}}^{t}_{\Lambda_{c}} + \mathcal{M^{\prime}}^{t}_{\Lambda^{\ast}_{c1}} + \mathcal{M^{\prime}}^{t}_{\Lambda^{\ast}_{c2}}
\end{eqnarray}
With the above total amplitude, one can obtain the differential  cross sections of the $\pi^{-} p \rightarrow D^{-} D^{0} p$ process,
\begin{eqnarray}
d{\sigma}=\frac{1} {8(2 \pi)^4} \frac{1} {	\Phi} \overline{\left| {\mathcal{M^{\prime}}_{\mathrm{Tot}}}\right|^2 }d p^0 _5 d p^0 _3 d\cos\theta d \eta,
\end{eqnarray}
where the flux factor $\Phi=4|{\vec{p_1}}|\sqrt{s}$, $\vec{p_1}$ and $\sqrt{s}$ represent the three momentum of the initial $\pi^{-}$ particle and the center of mass energy,  respectively. $p^{0}_{3}$ and $p^{0}_{5}$ are the energy of the outgoing $D^{-}$ meson and $p$, respectively. 

The cross sections for $\pi^{-} p \rightarrow D^{-} D^{0} p$ depending on the momentum of the incident pion beam with $\Lambda_r=1.1$ GeV are presented in Fig.~\ref{Fig:2}. Among them, Fig.~\ref{Fig:2}-(a), Fig.~\ref{Fig:2}-(b) and Fig.~\ref{Fig:2}-(c) correspond to the cross sections resulted from the $s$, $u$ and $t$-channel, respectively. In these three figures, the black solid curves represent the total contributions of each channels. The cyan dotted, red dashed and blue dash-dotted curves stand for the contributions from the processes that include propagators $\Lambda_{c}(2286)$, $\Lambda_{c}(2910)$ and $\Lambda_{c}(2940)$, respectively. As indicated in diagram $(a)$, the contributions from $\Lambda_c(2286)$ and $\Lambda_c(2940)$ are not shown in the diagram because their cross sections are less than 1fb. As for the $u$-channel, $\Lambda_c(2940)$ intermediate process is dominant, while the cross section of $\Lambda_c(2910)$ is about 0.1pb when $P_\pi=15$ GeV, which is several times smaller than that of $\Lambda_c(2940)$. But for the $t$-channel, the cross section resulted from $\Lambda_c(2910)$ intermediate process is predominantly, which is about 20 times of that from $\Lambda_c(2940)$, and about 450 times of that from $\Lambda_c(2286)$ when $P_\pi >13 $ GeV. By comparing Fig.~\ref{Fig:2}-(a)-(c), one can find the cross sections resulted from $s$, $u$, and $t$-channels are of orders of $10^{-2}~\mathrm{pb}$, $10^{-1}~\mathrm{pb}$, and $10^{2}~\mathrm{pb}$, respectively, which indicates that at high energy scattering process, the $t$-channel contributions are dominant. 

In Fig.~\ref{Fig:2}-(d), we present the individual contributions from $\Lambda_c(2286)$, $\Lambda_c(2910)$ and $\Lambda_c(2940)$, which correspond to cyan dotted, red dashed and blue dash-dotted curves, while the solid curve refers to the total cross sections for the $\pi^{-} p \rightarrow D^{-} D^{0} p$ process. From this diagram, one can find that the total cross sections increase rapidly near the threshold, and with $P_{\pi}$ greater than $12.5~\mathrm{GeV}$, the cross sections increase rather slowly with the increasing of $p_{\pi}$. At $P_{\pi}=15~\mathrm{GeV}$, the cross section reach up to 487 pb. In addition, the present estimations indicate that the dominant contributions to the cross sections for $\pi^{-} p \rightarrow D^{-} D^{0} p$ come from the intermediate state $\Lambda_c(2910)$.

In addition to the cross sections, we also investigate the differential cross section for $\pi^{-} p \rightarrow D^{-} D^{0} p$ depending on the invariant mass of $D^0 p$. The $D^0 p$ invariant mass distributions at $p_\pi =14$ GeV are presented in Fig.~\ref{Fig:4}. The cyan dotted, red dashed and blue dash-dotted  curves correspond to the individual contributions from $\Lambda_c(2286)$, $\Lambda_c(2910)$, and $\Lambda_c(2940)$, respectively, while the black solid curve refers to the summation of the individual contributions as well as their interferences. From this figure, one can find that the contributions from $\Lambda_c(2286)$ are rather smooth, which plays the role of the background. As for $\Lambda_c(2910)$ and $\Lambda_c(2940)$, our estimations suggested that the signal of $\Lambda_c(2910)$ is predominantly, while the signal of $\Lambda_c(2940)$ is about one order smaller than that of $\Lambda_c(2910)$, which indicate the expected structure near 2.9 GeV in the $D^0 p$ invariant mass spectrum of $\pi^{-} p \rightarrow D^{-} D^{0} p$ process should corresponds to $\Lambda_c(2910)$ rather than $\Lambda_c(2940)$.

It is worth noting that the cutoff parameter $\Lambda_{r}$ cannot be well determined experimentally at present. So in the present estimation, we vary $\Lambda_{r}$ from 1.0 to 1.2 $\mathrm{GeV}$ to further check the parameter dependence of the total cross sections for the $\pi^{-} p \rightarrow D^{-} D^{0} p$ process. Our estimated cross sections are presented in Fig.~\ref{Fig:3}, where the black curve is calculated with $\Lambda_{r}=1.1$ $\mathrm{GeV}$ and the cyan band indicates the uncertainties resulting of the parameter $\Lambda_{r}$. In the considered model parameter range, the total cross section at $p_\pi =15 $ GeV is estimated to be $(0.49^{+1.56}_{-0.38})$ nb at $p_{\pi}=15$ $\mathrm{GeV}$, which means that the values of the total cross sections crosse an order of magnitude when $\Lambda_{r}$ ranges from 1.0 to 1.2 $\mathrm{GeV}$.

Before the end of this work, it is worth mentioning that the cutoff employed in the present estimations is determined by the upper limit of the cross section for $\pi^- p \to D^{\ast -}\Lambda_c$ at $P_\pi =13$ GeV. Physically, the cutoff is related to the size of the interacting particles. Compared to the traditional $\Lambda_c$ baryon state, $\Lambda_c(2910)$ and $\Lambda_c(2940)$,  as the $ND^\ast$ molecular states, should exhibit a larger size. This suggests that the cutoff for the vertices involving $\Lambda_c(2910)/\Lambda_c(2940)$ should be smaller than 1 GeV, and leading to smaller cross sections.

\section{SUMMARY}
With the observations of $\Lambda_{c}(2910)$ and $\Lambda_{c}(2940)$ by the BABAR and Belle Collaborations, the investigations of these two  states have become intriguing. The observed masses of both $\Lambda_c(2910)$ and $\Lambda_c(2940)$ are close to the threshold of $D^\ast N$, this situation is similar to the well-known pentaquark candidates $P_c(4440)$ and $P_c(4457)$, and in addition, the mass splitting of $\Lambda_c(2940)$ and $\Lambda_c(2910)$ is very similar to that of $P_c(4440)$ and $P_c(4457)$. These particular spectrum properties inspire the $ND^\ast$ molecular interpretations to the $\Lambda_c(2910)$ and $\Lambda_c(2940)$.
 
Besides the mass spectrum and decay behaviors, the production properties of $\Lambda_c(2910)$ and $\Lambda_c(2940)$ can also shed light on the inner structure of these two states. In addition, the pion beam energy will reach over 20 GeV in the J-PARC hadron facility. Thus, in the present work, we propose to investigate the $\Lambda_c(2910)$ and $\Lambda_c(2940)$ productions in $\pi p$ scattering process. For parameter $\Lambda_{r}$, the measurement in Ref.~\cite{Christenson:1985ms} indicated that the upper limit of cross section for $\pi^{-} p \rightarrow D^{\ast-} \Lambda_{c}$ was 7 nb at $P_{\pi}=13$ GeV. Thus, we adopt $\Lambda_{r}=1.1$ GeV in the present estimation by comparing our estimations with the experimental measurement. The $\pi^{-} p \rightarrow D^{-} \Lambda_{c} / \Lambda_{c}(2910) / \Lambda_{c}(2940)$ processes are also estimated, and our estimations indicate that at $p_\pi=30$ GeV, the ratios of the cross sections for $\pi^- p \to D^{\ast-} \Lambda_c(2286)$, $\pi^- p \to D^{-} \Lambda_c(2286)$, $\pi^- p \to D^{-} \Lambda_c(2910)$, and $\pi^- p \to D^{-} \Lambda_c(2940)$ are $1:4.8:1.42:0.26$.

Considering the fact that $\Lambda_c(2910)$ and $\Lambda_c(2940)$ can decay into $D^0 p$, we propose to investigate these two states in the $D^{0}p$ invariant mass spectrum of $\pi^{-} p \rightarrow D^{-} D^{0} p$ process. Our estimations indicate that the cross sections for $\pi^{-} p \rightarrow D^{-} D^{0} p$ range from several hundreds pb to several nb with the varying of the model parameter $\Lambda_{r}$. Moreover, our estimations suggest that the dominant contributions come from $\Lambda_c(2910)$, and the expected structure around 2.9 GeV in the $D^0 p$ invariant mass spectrum of $\pi^{-} p \rightarrow D^{-} D^{0} p$ process should corresponds to $\Lambda_c(2910)$ rather than $\Lambda_c(2940)$, which could be tested by further experimental measurements at J-PARC in the future.

\section*{ACKNOWLEDGMENTS}
The authors would like to thank Prof. Ju-Jun Xie for useful discussions. This work is partly supported by the National Natural Science Foundation of China under the Grant Nos. 12175037 and 12335001, as well as supported, in part, by National Key Research and Development Program under the contract No. 2024YFA1610503

 
\bibliographystyle{unsrt}
\bibliography{ref.bib}
\end{document}